\documentclass[a4paper,11pt]{article}
\usepackage{lmodern}
\usepackage{graphicx}
\usepackage{amssymb} %
\usepackage{amsmath}
\usepackage{makeidx}     
\makeindex 
\newtheorem{thm}{Theorem}[section]

\newtheorem{proposition}[thm]{Proposition}

\begin{document}
\title{\Large{\textbf{Analysis of an M/G/1 system for the optimization of the RTG performances in the delivery of containers in Abidjan Terminal}}}
\date{}
\maketitle  
\begin{center}
	\author \footnotesize{Bakary Kon\'e$^{1}$, Salimata Gueye Diagne$^{1}$,  D\'ethi\'e Dione$^{1}$, Coumba Diallo$^{1}$}\\
          bakary.kone@ucad.edu.sn\\
          dethiedione79@gmail.com
\end{center}  

	\begin{center}
		\scriptsize{1-Universit\'e Cheikh Anta Diop de Dakar, Senegal}
	\end{center}  
\begin{abstract}
\textsl{In front of major challenge to increase its productivity while satisfying its customer, it is today important to establish in advance the operational performances of the RTG Abidjan Terminal. In this article, by using an M/G/1 retrial queue system, we obtained the average number of parked delivery trucks and as well as their waiting  time. Finally, we used Matlab to represent them graphically then analyze the RTG performances according to the traffic rate.}  

\end{abstract}
\textbf{Key words}\\
 \scriptsize{ RTG, Embedded Markov chain, Retrial Queue, Stationary Distribution, Performance Measures}
\normalsize
\section{Introduction}
In today's maritime world, marked by increasing modernity, containers terminals do not only have the obligation to straighten their infrastructures, but also they new ones have to improve to keep their customer and acquire. This task is not easy given the deep international competition. A task Terminals with containers show awareness by trying to satisfy ever increasing their customer.
This report was made at Container Terminal of Abidjan (Abidjan Terminal), where one of the priorities is to satisfy the customer in particular trucks in mission of delivery to the city \cite{abd}.
A vast distribution platform of containers of any singles and the modern installations, allow Abidjan Terminal to receive and to handle a high volume of imported containers and a more and more demanding customer.
Given its important role in the Ivory Coast maritime exchanges, it is important to analyze the RTG performances which establish its most used porticoes of courses in the operations of containers deliveries \cite{abn}.\\
By using the queue theories, several researchers have already carried out studies which analyses the performances in various domains. 
It is the case of Mohamed Boualem and al. $(2013)$ who deal retrial queue system and Bernoulli feedback \cite{Bbb}. They considered the type $M/G/1$ retrial queue and feedback system. By using the same analysis, they obtained important performances measures.
The researchers Yang and Templeton $(1987)$, as well as Yang and al. $(1994)$ also used the  $M/G/1$ retrial model in the case of exponential and general distributions of the inter-retrial time. They proved that the generative function of the number of customers in the $M/G/1$ retrial system is the product of two generative functions: that of the number of customers in the $M/G/1$ classic system\cite{Art} and the generative function for the number of customers in the $M/G/1$ retrial system given that the server is free \cite{Btt}.\\
As for Stidham $(2002)$ another important class of queue deals the systems in which the server remains idle when the queue is empty. So, this time of idleness of the server could be used differently with the aim of improving the efficiency of the system \cite{Mmm}.\\
Aissani $(2008)$ made a complete analysis of the constant retrials in the case of the Quasi-Markovian  $M/G/1$ models. He considers a queue of this type with the politics of constant retrial and policy vacancy of the server, when the retrial time, the service time and the policy vacancy time are arbitrarily distributed. The distribution of the number of customers in the stationary system mode is obtained in terms of generative function. Afterward, he gives an approximation for such a distribution\cite{Alf}.\\
Researchers Djamil Aissani $(2011)$ consider the example of the $M/G/1$ classic retrials queue and server vacancies. By taking into account the existence of the stationary mode and an embedded Markov chain  they make a stationary analysis of the system. They also obtained formula for the limit distribution of the server state, the stochastic decomposition\cite{Anb} and some performance measures \cite{Hyy}.\\
In this paper, we use the theories of retrials queues to clear and analyze the RTG performances in Abidjan Terminal. This question was not approached in the literature, which is the main contribution of this article.\\
We consider trucks parked in the zone under RTG and their treatment ordered as a Quasi-Markovian $M/G/1$ queue system with a period of the RTG vacancy. This analysis allows to establish in advance the operational performances of porticoes, to identify the critical elements or, again, to look at the effects of a modification of the operating conditions. We use the method of the generative functions, then present a more detailed analysis with the calculation of the generative functions.\\
The rest of the document is organized as follows: the second section deals with the mathematical description of the model. In the third section, we present an embedded Markov chain and an existence condition of the stationary state of the embedded Markov chain. In the fourth section, we release some clear performance measures. That is going to allow us to release in the fourth section a digital test, and finally a conclusion and perspectives are proposed in the last section.\\
\section{Delivery and operational function of the RTG}
As well as its role of storage space, the zone under RTG constantly receives containers in import, export and transshipment. But in this section we will present the process of containers delivery.
\begin{figure}[!h]
	\centering
	\includegraphics[width=0.70\textwidth]{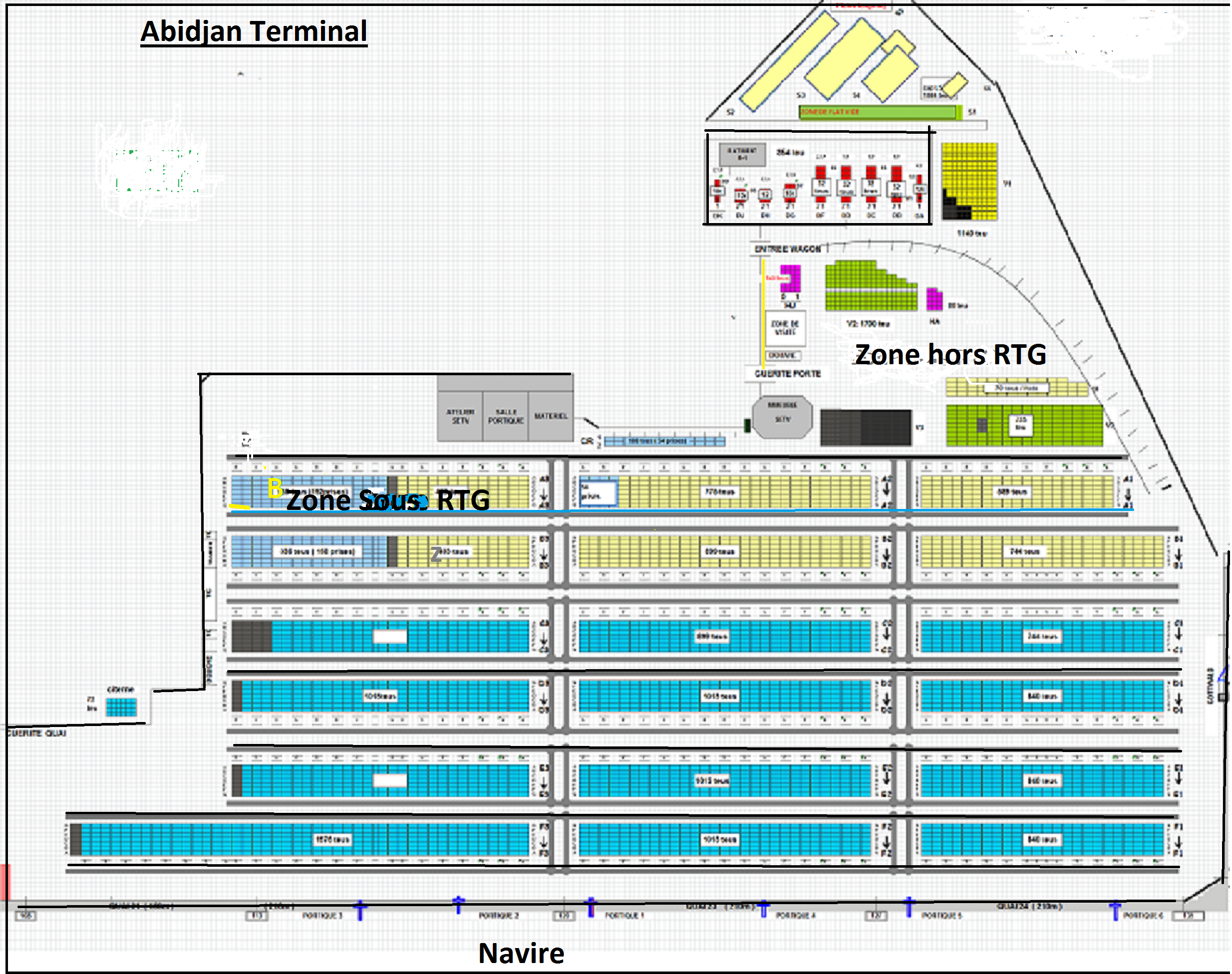}
	\caption{Map of Abidjan Terminal}
	\label{fig1}
\end{figure}
\\
$\blacklozenge$ \textbf{Delevery process}\\
\\
The driver arrives at the sentry box of the Terminal provided with a delivery slip.
After routine check of the number of the container, the sentry agent affects on Oscar (Operating Software Container and adjournment is a management system of exploitation of the operations of a Container terminal) a token with a number to this container then confirms (putting in delivery). Then he returns the token to the driver. The latter leads the truck on the park and parks at the address of the container indicated on the token.
The RTG operator having received the mission loads the truck after checking its numbers (container and the token) (mission in processes). The loaded truck returns to the exit sentry box for the end of the delivery procedure.\\
\\
$\blacklozenge$ \textbf{The RTG operational function}\\
\\
The RTG has a mission of execution which handling operations are planned and checked since the control room by the piloting. Indeed, according to the number of containers to be delivered and of their localizations the import piloting proceeds to a division of the zone under RTG on the graphic software OSCAR in small zones (working zone) corresponding to the available RTG. Then, he allocates to every working zone a RTG which receives and executes the missions in its zone. Finally the piloting follows and coordinates the movements of the RTG in the execution of the various operations of the RTG since the control room.
\section{Research Question}
The zone under RTG which constitutes our workspace is formed by $18$ blocks parallel to the linear band of the Terminal Quays and the containing each several bays of various types of containers.
Containers are piled in standard ISO there, according to their statuses at using addresses to allow their localizations.
During the delivery phase of containers, trucks make their entrance by the procedure and park at this address of the container.
The RTG operator who covers the zone receives immediately the number of the token from a small screen placed in his cabin.
This list reached him in arrival order of trucks. So, after checking the information, (the numbers of the container and that of the token) the RTG operator loads the container.
Contrary to arrival spontaneous of the trucks, the RTG being in lower number cannot handle all the containers at reasonable time.
The loading is done at a disproportional rhythm in the flows of parked trucks.
In order to increase the operational and competitive qualities of the Abidjan Terminal, it is important to analyze the performance of its machines to make it more attractive. 
In this article, by using the generative functions, we analyze the RTG performances which allows to identify the critical elements or, again, to grasp the effects of a modification of the operating conditions.

\section{Modeling problem}
It is composed of several stages:
\subsection{Model description}
Delivery trucks make their orderly entry and park at the precise addresses of containers. The zone can contain several trucks which makes its structure look like a classic queue where the server is the RTG. The parked trucks are the considered first category or the second category customers according to their waiting times. While a truck parks and is not served for the moment the service of piloting makes retrials to the RTG operator so that this truck is loaded. Once served, it immediately leaves the Terminal with the container.\\
$\bullet$ The primary trucks are those which are immediately loaded upon their arrival.\\
$\bullet$ The secondary trucks are the ones which wait for a whole. These trucks are loaded further to one or several retrials made by the piloting.
Indeed, to avoid of lengthy waiting, the piloting follows and coordinates the movements of machines on the Park. He constantly reminds the RTG operators to handle first and foremost the parked trucks having spent a long time waiting. 
This principle of reminder is defined according to the number of parked trucks and the duration of their waiting time. That increases in intensity when the zone of the RTG is saturated.
The intervals the retrials time are they follow an exponential distribution rate $k\theta$ where ($k \in N$ is the number of waiting trucks and $\theta > 0$ is the debit of load of a truck by reminder) (Boualem and $al.2013$). All the trucks entering the system are served in a random order independent from their arrival time. The service time distributed according to an any distribution which the density $b(t)$ is transformed by Laplace-Stieltjes $\overline{B}(t)$ for $t$ at a given moment.
That means of $\tau_{i}^{s}$ is the service which is between the $i^{th}$ and $(i+1)^{th}$ service, then P $(\tau_{i}^{s}\leq t) = B(t)$. If we are in the $i^{th}$ stage, the RTG executes the $i^{th}$ service there which corresponds to the loading of the $i^{th}$ delivery truck. The time of $(i+1)^{th} $ begins just at the end of the loading of the $i^{th}$ truck. It's established by the travel time made by the RTG between these two trucks and by the unproductive movements time as well as the loading of the container. The first two moments finished of this distribution are $\beta_1$ and $\beta_2$, respectively. We suppose that the RTG is free every time the system becomes empty(with no trucks in the zone)\\
Either $t$ represents any moment. The state of the system at this arbitrary moment is given by:\\
$$M(t)=\{C(t), N(t), \zeta (t),\}$$\\
With,
$$C(t)=\left\{\begin{array}{l l}
 1& \text{If the RTG is engaged} \\
 0& \text{Otherwise}
 \end{array}\right.$$
 $N(t)$ represents the number of trucks parked at a given moment $t$\\
 $\tau(t)$ represents the duration of the RTG service to handle the truck if $C(t)=1$.
\subsection{Embedded Markov chain}
The Poisson distribution of an $M/G/1$ queue which sequences can be described there
terms of a sequence alternated by various periods. So, the RTG works in a continuous way and only becomes free when all the parked trucks have been served. It resumes the handling when a truck arrives. \\
We can summarize these parameters in the table below.
\begin{table}[h!]
	\centering
	\caption{Parameter setting}
	\label{tab:3}       
	\begin{tabular}{|p{2cm}||p{9.5cm}|}
		\hline
		\centering Parameters &  Definitions \\
		\hline
		\centering $i$ & The stage in which a service was made by the RTG. A service corresponds to a stage which amounts to a truck loaded.\\
		\hline
		\centering $k, j$ & The number of trucks in waiting.\\
		\hline
		\centering $\zeta_{i}$ & The exact time from which the $i^{th}$ service is ended.\\
		\hline
		\centering $\tau^{s}_{i}$ & The elapsed time by the RTG for the $i^{th} $ service \\
		\hline
		\centering $N(\zeta_{i})$ & The total number of waiting trucks parked after the $i^{th}$ service. \\
		\hline
	\end{tabular}
\end{table}

Let us consider $\{\zeta_{i}, \; i\in \mathcal{N} \} $ an increasing continuation of moment of the completion of a service time.
Or $N(\zeta_{i})$ the number of trucks parked waiting in the system just after the loading of the $i^{th} $ truck. 
We then define a stochastic process by the sequence of the random variables:
 $$Q_{i} = \{N (\zeta_{i}); \; i=0, 1, 2, 3, \cdots \} $$ 
 where $\zeta_i$ is the exact moment of the end of the treatment of the $i^{th}$ truck. 
\\ 
Let us pose: $$N (\zeta_i) N_{i} $$
And $A_{i+1}$ the number of parked trucks while ${(i+1)}^{th}$ truck is being served. Then, the random variables $A_{i}$ are independent between them; their common distribution is given by:
\begin{equation}
P(A_{i+1}=k)=q_{k}=\int_{0}^{+\infty}\frac{(\lambda t)^k}{k!}e^{-\lambda t}b(t)dt; \; q_{k}\geq 1,\;(k=1, 2,...)
\end{equation}
Its generative function is:
\begin{align*}
A(z)&=\sum\limits_{k=0}^{+\infty}q_{k}z^{k}\\
&=\overline{B}(\lambda (1-z))
\end{align*} 
\underline{The fundamental equation of embedded Markov chain}\\
\\
When the RTG is handling $(i+1)^{th}$ truck, if $A_{i+1} $ arrive, then at the end of his load we shall have:
$$N_{i+1}=\left\{\begin{array}{l l}
N_{i}-1+ A_{i+1}$\; si $ \; N_{i}\geq 1 \\
A_{i+1}$\; si $\; N_{i}=0
\end{array}\right.$$ 
Hence the fundamental equation of embedded Markov chain is given by:\\
\begin{equation}
N_{i+1}=N_{i}-\delta_{N_i}+A_{i+1}
\end{equation}
The variable of Bernoulli $\delta_{N_i}$ is defined by:\\
$\delta_{N_i}=\left\{\begin{array}{l l}
1 & \quad \text{ If the $(i+1)^{th}$ Served truck is from the second category }\\
0 & \quad \text{Otherwise}
\end{array}\right.$ 
depend on ${N_{i}}$. It's translated by in most $1$ truck left and $A_{i+1 }$ trucks arrived at the end of $(i+1)^{th} $ service.\\
\\
\underline{Conditional distribution} \\
\\
For $k$ trucks waiting in the stage $i$, the following state is determined according to the value taken by $\delta_{N_i}$. Then the conditional distribution is given by (Boulem $2011$)
\begin{equation}
P(\delta_{N_i}=1/N_{i}=k)=\frac{k\theta}{\lambda +k\theta}$$
$$P(\delta_{N_i}=0/N_{i}=k)=\frac{\lambda}{\lambda +k\theta}
\end{equation}
With $\theta$ the flow of truck loaded from the second category.\\
\\
\underline{Transition rate}.\\
\\
The probability that there is $j$ trucks parked in $(i+1)^{th}$ stage knowing that it had $k$ the $i^{th}$ is noted:\\
$\forall\; j\geq 0$ and $0\leq k\leq j$, we have:
\begin{equation}
P_{kj}=P(N_{i+1}=j/N_{i}=k)
\end{equation}
Which gives:\\ 
$$P_{kj}=P(A_{i+1}=j-k)P(\delta_{N_i}=1/N_{i}=k)+P(A_{i+1}=j-k+1)P(\delta_{N_i}=0/N_{i}=k)$$
According to $(1)$ and $(4)$ we have:\\
\begin{equation}
P_{kj}=\left\{\begin{array}{l l}
P(N_{i+1}=j/N_{i}=k)=q_{j-k}\frac{k\theta}{\lambda +k\theta}+q_{j-k+1}\frac{\lambda}{\lambda +k\theta}, \quad 0\leq k\leq j\\
0\quad \text{Otherwise}
\end{array}\right. 
\end{equation}

We remind that $\zeta_i$ is the precise moment in which the treatment of the $i^{th} $ truck parked in the line ended. \\
We verify that this continuation of random variable is an actually Markov chain in discrete time. \\ 
Yet $A_{i}$ the number of parked trucks while the $i^{th} $ truck is being served.\\
 \\
\underline{Existence condition of the stationary distribution of the number of trucks}\\
\\
The chain is Markovian because the state of the system in the stage $(i+1)$, depends on the state $i$. So that the line does not lengthen infinitely, the average number of arrivals during the duration of a service must be strictly lower than $1$. Then, we have:\\
\begin{align*}
E(A_i)&=\sum\limits_{k=0}^{i}\int_{0}^{+\infty}P(A_{i}=k)b(t)dt\\
    &=\sum\limits_{k=0}^{i}k\int_{0}^{+\infty}\frac{(\lambda t)^k}{k!}e^{-\lambda t}b(t)dt\\
   &=\int_{0}^{+\infty}e^{-\lambda t}b(t)\lambda t\{\sum\limits_{k=0}^{+\infty}\frac{(\lambda t)^i}{k!}\}dt\\
 &=\lambda \int_{0}^{+\infty}tb(t)dt\\ 
 &=\lambda \beta_{1}
\end{align*}
With $\beta_{1}$ the moment of order $1$ of the distribution. 
Hence the existence of the stationary distribution is conditioned for:
 \begin{equation}
\rho =\lambda\beta_1 \leq 1	
\end{equation}
\underline{The generative function of the trucks number waiting in the RTG zone}\\

It corresponds to the generative function of $N_{i}$ and thus is the transformed $z$ of its function of density which is defined by:
$$f_{i}(z)=E(z^{k})=\sum\limits_{k=0}^{+\infty}z^{k}P[N_{i}=k]$$	
Yet $$N_{i+1}=N_{i}+\delta_{N_i}+A_{i+1}$$
$$f_{i+1}=E[z^{N_{i}-\delta_{N_i}+A_{i+1}}]$$
As $A_{i}$ and $N_{i}$ are independent then, 
\begin{align*}
f_{i+1}&=E[z^{N_{i}-\delta_{N_i}+A_{i+1}}]\\
&=E[z^{N_{i}-\delta_{N_i}}]E[z^{A_{i+1}}]\\
\end{align*}	
Given that the condition of stationarity is then verified $A(z) = E(z^{A_{i}})$ independent of $i$. Thus we have:
\begin{align*}
f_{i+1}&=E[z^{N_{i}-\delta_{N_i}}]E[z^{A_{i+1}}]\\
&=E[z^{N_{i}-\delta_{N_i}}]A(z)\\
&=[\sum\limits_{k=0}^{+\infty}z^{k-\delta_{k}}P(N_{i}=k)]A(z)\\
&=[\sum\limits_{k=0}^{+\infty}z^{k}P(N_{i}=k)\times P(\delta_{N_i}=0/N_{i}=k)\\
&+\sum\limits_{k=1}^{+\infty}z^{k-1}P(N_{i}=k)\times P(\delta_{N_i}=1/N_{i}=k)]A(z)\\
&=\sum\limits_{k=0}^{+\infty}z^{k}P(N_{i}=k)\frac{\lambda}{\lambda+k\theta}+\sum\limits_{k=1}^{+\infty}z^{k-1}P(N_{i}=k)\frac{k\theta}{\lambda +k\theta}\\
\end{align*}
In the stationary state, \\
$\lim\limits_{i\longmapsto +\infty}f_{i+1}(z)=\lim\limits_{i\longmapsto +\infty}f_{i}(z)=f(z)$ and the stationary distribution as defined by:\quad $\pi_{k}=\lim\limits_{i\longmapsto +\infty}P(N_{i}=k)$
By using them in the equation, we obtain:
\begin{equation}
f(z)=\sum\limits_{k=0}^{+\infty}z^{k}\pi_{k}\frac{\lambda}{\lambda+k\theta}+\sum\limits_{k=1}^{+\infty}z^{k-1}\pi_{k}\frac{k\theta}{\lambda +k\theta}
\end{equation}
We pose: $$\psi(z)=\sum\limits_{k=0}^{+\infty}\frac{\pi_{k} z^{k}}{\lambda+k\theta} $$
Then 
$$\psi'(z)=\sum\limits_{k=1}^{+\infty}\frac{k\pi_j z^{k-1}}{\lambda+k\theta}$$\\
So, the equation $(10)$ allows us to obtain the following relation:
\begin{equation}
f(z)=A(z)[\lambda\psi(z)+\theta\psi'(z)]
\end{equation}
In a different way, we notice that:\\
\begin{align*}
f(z)&=\sum\limits_{k=0}^{+\infty}\pi_k z^{k}\\
&=\sum\limits_{k=0}^{+\infty}\pi_{k} z^{k}\frac{\lambda+k\theta}{\lambda+k\theta}\\
&=\lambda\psi(z)+\theta\psi'(z)
\end{align*}
Hence the relation:\\
\begin{equation}
f(z)=\lambda\psi(z)+\theta\psi'(z)
\end{equation}\\
Consequently, the use of the equations $(11)$ and $(12)$ entails:\\
$$
\lambda\psi(z)+\theta\psi'(z)=A(z)[\lambda\psi(z)+\theta\psi'(z)]$$
\begin{equation}
\theta\psi'(z)[A(z)-z]=\lambda\psi(z)[1-A(z)]
\end{equation}
The function $h(z)=A(z)-z$  positive, increasing for $z\in [0, 1]$ and $\rho<1$, $z<A(z)<1$ \cite{BTT} so $h(1)=1$ \\
$h'(z)=A'(z)-z'$, and $h'(1)=\rho-1<0$. 
 Besides by using the results $(4)$ and $(5)$, we have:\\
\\
 $$\psi'(z)=\frac{\lambda}{\theta}\{\frac{1-A(u)}{A(u)-u}\}\psi(z)$$\\
 For $\rho<1$, The resolution of this differential equation gives us:\\
  \begin{equation}
\psi(z)=\psi(1)exp{\{\frac{\lambda}{\theta}\int_{1}^{z}\frac{(1-A(u))}{A(u)-u}d_u\}}
 \end{equation}
 Yet $$f(z)=\lambda\psi(z)+\theta\psi'(z)$$\\
Therefore
 \begin{align*}
  f(z)&=\lambda\psi(z)+\{\theta \frac{\lambda}{\theta}z\frac{(1-A(z))}{A(z)-z}\}\psi(z)\\
  &=\lambda\psi(z)A(z)\frac{1-z}{A(z)-z}
 \end{align*}
The traffic intensity is $\rho=\lambda\beta_{1}$ and 
 $f(1)=\overline{B}(0)=1$. So a simple calculation gives us  $\frac{1-z}{A(z)-z}=h'(1)=\rho-1$ then  $\psi(1)=\frac{1-\rho}{\lambda}$.\\
By replacing $ \psi (1) $ by its expression we obtain the generative function of the stationary distribution of the number of trucks waiting is:\\ 
\begin{equation}
f(z)=\frac{(1-\rho)\overline{B}(\lambda-\lambda z)(1-z)}{\overline{B}(\lambda-\lambda z)-z}\times exp{\{\frac{\lambda}{\theta}\int_{1}^{z}\frac{[(1-A(u))]}{A(u)-u}d_u\}}
\end{equation}
\section{The RTG performance measures } 
In this section, we establish the explicit expressions of the RTG performance measures during the operations of delivery of containers in this form.\\
\begin{proposition}
In the stationary state, the average number of trucks in waiting in the RTG zone and the average  waiting time of the delivery trucks of containers are respectively expressed by: 
\begin{equation}
\overline{N}=\rho+\frac{\lambda\rho}{\theta(1-\rho)}+\frac{\lambda^{2}\beta_{2}}{2(1-\rho)}
\end{equation}
\begin{equation}
\overline{W}=\beta_{1}+\frac{\lambda \beta_{2}}{2(1-\rho)}+\frac{\rho}{\theta(1-\rho)}
\end{equation}
\end{proposition}
\textbf{Demonstration}\\
$\bullet$ Let us determine the average number of trucks waiting in the RTG zone. \\ 
In the stationary state, the average number of trucks being in the RTG zone corresponds in numbers of trucks which were already parked and those which have just arrived. It thus expresses itself by:
$$\overline{N}=\lim\limits_{i\longmapsto +\infty}E(N_{i+1})=\lim\limits_{i\longmapsto +\infty}E(N_{i})$$\\
Because $f (z)$ of its generative function, 
This allows to say that $$\overline{N}=f'(1)$$ 
With $f'(z)$ the derivative of the generative function.\\
So we calculate $f(z)$, then we replace $z$ by $(1)$ to obtain the different performances.\\ 
\\
\underline{Calculation of the derivative of the generative function}\\
$$f(z)=\frac{(1-\rho)\overline{B}(\lambda-\lambda z)(1-z)}{\overline{B}(\lambda-\lambda z)-z}\times exp{\{\frac{\lambda}{\theta}\int_{1}^{z}\frac{[(1-A(u))]}{A(u)-u}d_u\}}$$
We pose:\\
$G(u)=\frac{[(1-A(u))]}{A(u)-u}$, \quad where \quad $1\leq u\leq z$\\
$T(z)=(1-\rho)\overline{B}(\lambda-\lambda z)(1-z)$;\quad $P(z)=exp\{\frac{\lambda}{\theta}\int_{1}^{z}g(u)du\}$ and \\$Q(z)=\overline{B}(\lambda-\lambda z)-z$\\So\\
\begin{equation}
T(1)=0,\quad P(1)=1,\quad Q(1)=0
\end{equation}
$$T'(z)=(1-\rho)\overline{B}^{1}(\lambda-\lambda z)(1-z)+(\rho-1)\overline{B}(\lambda-\lambda z)$$
$$P'(z)=\frac{\lambda}{\theta}\times G(z)exp\{\frac{\lambda}{\theta}\int_{1}^{z}g(u)du\}$$ 
Then
\begin{equation}
T'(1)=(\rho-1),\quad 
P'(1)=-\frac{\lambda}{\theta}\times\frac{\rho}{1-\rho},\quad Q'(1)=(\rho -1)
\end{equation}
Let us now notice that:
$$f(z)=\frac{T(z)P(z)}{Q(z)}$$
Because we have: $$\frac{1-z}{\overline{B}(\lambda-\lambda z)-z}=h'(1)=1$$ 
Then
\begin{equation}
f(1)=1
\end{equation}
Therefore
$$f(z)Q(z)=T(z)P(z)$$
$$f'(z)Q(z)+f(z)Q'(z)=T'(z)P(z)+T(z)P'(z)$$
By replacing each function by its expression 
$Z=1$, we have an indefinite shape. We raise it by the application of the hospital theorem.\\ 
The second derivative gives:\\
$$f''(z)Q(z)+2f'(z)Q'(z)+f(z)Q''(z)=T''(z)P(z)+2T'(z)P'(z)+T(z)P''(z)$$
The equations $(13)$ allow to obtain:\\
\begin{equation}
f'(z)=\frac{T''(z)P(z)+2T'(z)P'(z)-f(z)Q''(z)}{2Q'(z)}
\end{equation}
Yet $$T''(z)=2\rho(\rho-1)$$
By using the formula of the moments:
$\beta_{k}=(-1)^{k}\beta^{k}(0)$\\
$$\beta_{1}=-\beta^{1}(0),\quad
\beta_{2}=\beta^{2}(0)$$
The equations $(14)$ and $(15)$ entail:
\begin{equation}
f'(1)=\frac{2\rho(\rho-1)+2\frac{(\rho-1)\times\lambda\times \rho}{\theta\times(1-\rho)}-\lambda^{2}\beta_2}{2(\rho-1)}
\end{equation}
In conclusion,\\
$\bullet$ The average number of trucks waiting in the zone of the RTG is given by:
$$\overline{N}=\rho+\frac{\lambda\rho}{\theta(1-\rho)}+\frac{\lambda^{2}\beta_{2}}{2(1-\rho)}$$
This expression is the total sum of the trucks which were parked just at the beginning and those which arrived during $(i+1)^{th}$ service, the average number of trucks which parks during the duration of its service is $\rho$. \\
$\bullet$ Let us determine the average waiting time of the delivery trucks of containers\\
Let us apply the formula of \underline {Little}. So this number expresses itself by:\\
$$\overline{W}=\frac{\overline{N}}{\lambda}$$
Which gives:
$$\overline{W}=\beta_{1}+\frac{\lambda \beta_{2}}{2(1-\rho)}+\frac{\rho}{\theta(1-\rho)}$$
\section{Simulation and Analysis of the results}
To simulate our results, we limit our work to an RTG zone. We analyze the influence of the traffic rate on the stationary characteristics of the system to obtain a tendency of the variations of the number of trucks and their waiting time. For that purpose, we consider the variation of the traffic in the $M/G/1$ system.
During a service we have one $(1)$ truck loaded hence the service level is $\frac{1}{\beta_1}$ 1. So, the rate of retrials is $1,4$ and corresponds to the retrial average made by the piloting \cite{pro}.\\
By using the Matlab software we obtain the influence of the traffic rate on the numbers of waiting trucks and their waiting times. The obtained results are represented in a form of curved graphs below.\\

\begin{figure}[!h]
\centering
\includegraphics[height=4cm,width=10cm]{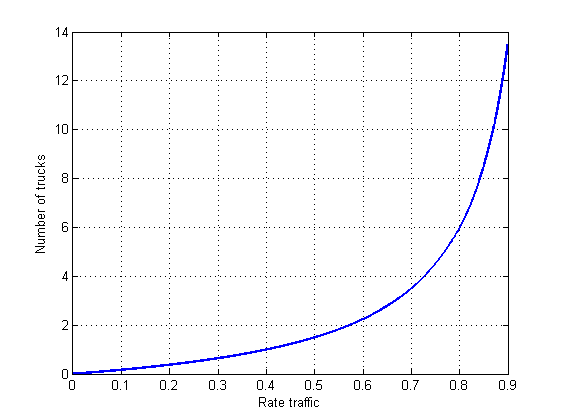}
\caption{Evolution of the trucks number in the system}
\label{fig18}
\end{figure}
We notice that there is a big dependency between the variations of the performance measures and the volume of traffic as shown in graph \ref {fig18}. This dependency is very sensitive for both cases of performances which we studied. The waiting time of trucks increases gradually according to the traffic, but as for the number of trucks, it evolves in an exponential way from $0.6$ arrival. For that purpose, you only have to increase the fluidity of the traffic so that the number of trucks in the system or the waiting time increases. Depending on the arrival flow, the number of trucks increases. The sensibility is more noticed when the arrival rate is close to one $(1)$. \\

\begin{figure}[!h]
	\centering
	\includegraphics[height=4cm,width=10cm]{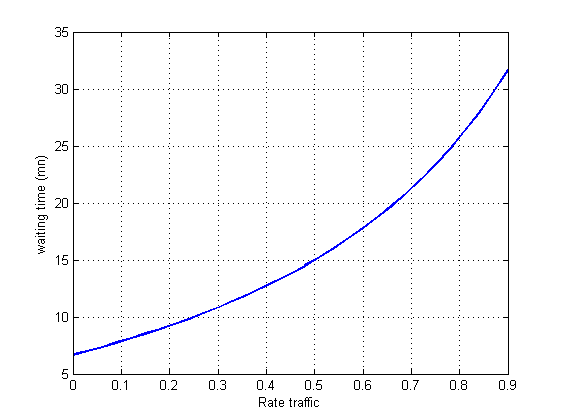}		
	\caption{Evolution of time waiting in the system}
	\label{fig20}
\end{figure} 

We can say that the unproductive movements contribute to the slowing down of the operational flow of the Abidjan Terminal a lot. The RTG spend a lot of time in the moving and the removing of containers. The numbers of parked trucks and the waiting time when they spend the park depend considerably on the traffic rate. The management of these two fundamental characteristics depends on the evolution of the traffic rate. The  higher it is, the longer waiting time and delivery trucks spend on the park as indicates in figure \ref{fig20}.
\section{Conclusion}  
In this article, we made our works on the operation zone of an RTG in Abidjan Terminal. By using the method of the generative functions, the theory of quasi Markovian retrial queue allowed us to obtain the performances such as the number of trucks and their waiting time. We used the software of programming $13.10$ version of the Matlab for graphical representations to analyze and release the sensitivity of the operations processing regarding to delivery trucks in the Terminal. The obtained results allowed us to say that the waiting times of delivery trucks depend widely on the traffic rate. The operational performances of the RTG depend widely on the traffic rate. \\
In perspective, we plan to estimate the unproductive movements then the probability of breakdowns of the RTG by using the same factor of sensitivity.\\
\textbf{Acknowledgements}\\
\textit
 {I thank the Director of the Abidjan Terminal for welcoming me as a trainee. I am very grateful to Mrs Asta Rosa CISSE for her moral support.\\
 I am extremely gratitude to Mr Laurent Bodji KASSI Chief Operating Officer and all his team for their simplicity, their loyalty and their professional quality}.
\normalsize
\bibliographystyle{mdpi}

\begin{thebibliography}{999}
\bibitem{Hyy}
Choo. H. Q and Conolly. B $(1979)$. New results in the theory of repeated orders queueing syst\`ems.\;{\em Journal of applied probability} $16: 631-640$
\bibitem{BTT}
Jackson J. R $(1961)$, waiting time distribution for Queues with dynamic priorities  {\em Naval Research logistics quarterly}, $9: 31-68$
\bibitem{Btt}
Jackson J. R $(1962)$, Queues with dynamic priorities  {\em Management science}, $1: 18-34$
\bibitem{Mmm}
Mohamed B. Natalia. D. Djamil A. $(2011)$,Approche r\'eg\'en\'erative de la file d'attente
M/G/1 avec rappels classiques et vacances
exhaustives du serveur. {\em Journal europ\'een des syst\`emes automatis\'e}, $45: 253-267$
\bibitem{Bbb}
Natalia. D. Djamil A. Mohamed B. $(2013)$, Analyse des performances du syst\`eme 
M/G/1 avec rappels et Bernoulli feedback. {\em Journal europ\'een des syst\`emes automatis\'e}, $47: 181-193$
\bibitem{Bbb}
ARRAR Nawel Khadidja, $(2012)$ Probl\`emes de convergence, optimisation d'algorithmes et
analyse stochastique de syst\`emes de files d'attente avec rappels. {\em https://tel.archives-ouvertes.fr/tel-00829089}, PP: $35-51$
\bibitem{Alf}
K. Jeganathan, Finite source Markovian inventory system with bonus service for certain customers, Int. J. of Adv. in Aply. Math. and Mech.  2(3) (2015) 134 - 143.
\bibitem{Anb} 
N. Anbazhagan, B. Vigneshwaran, K. Jeganathan, Stochastic inventory system with two types of services, Int. J. of Adv. in Aply. Math. and Mech. 2(1) (2014) 120 - 127.
\bibitem{Art} 
P. K. Pandit, Supply-and-demand model involving fuzzy parameters, Int. J. of Adv. in Aply. Math. and Mech. 1(2) (2013) 103 - 115.
\bibitem{pro} 
www.Abidjan terminal.com//content/view/69/85/lang,fr/
\bibitem{abn}
www.Abidjan terminal.com//content/view/12/13/
lang,fr/
\bibitem{abd}
www.Abidjan terminal.com//content/view/48/60/
lang,fr/
\end{thebibliography}
\makeatletter
\renewcommand\@biblabel[1]{#1. }
\makeatother

\end{document}